\documentclass[twoside]{article}
\usepackage{amsmath}
\usepackage[psamsfonts]{amssymb}
\usepackage{cmmib57}
\usepackage{diagrams}
\newcommand{\bPf}{\par\vspace*{-4pt}\indent{\sc Proof.}\enskip}
\newcommand{\ePf}{\medskip}
\def\QED{\hskip0.1em\hfill\null\ \null\nobreak\hfill\kern3pt\vbox{\hrule\hbox
   {\vrule\kern1pt\vbox{\kern1.7pt\hbox{$\scriptscriptstyle{QED}$}
    \kern0.2pt}\kern1pt\vrule}\hrule}}

\def\END{\hskip0.1em\hfill\null\ \null\nobreak\hfill\kern3pt\vbox{\hrule\hbox
   {\vrule\kern1pt\vbox{\kern1.7pt\hbox{$\,\,\,\vspace{5pt}$}
    \kern0.2pt}\kern1pt\vrule}\hrule}}
\newtheorem{theorem}{Theorem}
\newtheorem{lemma}{Lemma}
\newtheorem{corollary}{Corollary}
\newtheorem{proposition}{Proposition}
\newtheorem{remark}{Remark}
\newtheorem{definition}{Definition}
\newtheorem{example}{Example}
\newcommand{\bCd}{\bEq\begin{CD}}
\newcommand{\eCd}{\end{CD}\eEq}
\newcommand{\bcd}{\beq\begin{CD}}
\newcommand{\ecd}{\end{CD}\eeq}
\newcommand{\ben}{\begin{enumerate}}
\newcommand{\een}{\end{enumerate}}
\newcommand{\bEq}{\begin{eqnarray}}
\newcommand{\eEq}{\end{eqnarray}}
\newcommand{\beq}{\begin{eqnarray*}}
\newcommand{\eeq}{\end{eqnarray*}}
\newcommand{\bDf}{\begin{definition}\em}
\newcommand{\eDf}{\end{definition}}
\newcommand{\bLm}{\begin{lemma}}
\newcommand{\eLm}{\end{lemma}}
\newcommand{\bPr}{\begin{proposition}}
\newcommand{\ePr}{\end{proposition}}
\newcommand{\bTh}{\begin{theorem}}
\newcommand{\eTh}{\end{theorem}}
\newcommand{\bCr}{\begin{corollary}}
\newcommand{\eCr}{\end{corollary}}
\newcommand{\bRm}{\begin{remark}\em}
\newcommand{\eRm}{\end{remark}}
\newcommand{\bEx}{\begin{example}\em}
\newcommand{\eEx}{\end{example}}
\newcommand{\ie}{{\em i.e$.$} }
\newcommand{\eg}{{\em e.g$.$} }
\newcommand{\R}{I\!\!R}

\newcommand{\mto}{\mapsto}

\newcommand{\der}{\partial}

\DeclareMathOperator{\im}{im}
\DeclareMathOperator{\byd}{{\raisebox{.1ex}{:}{=}}}
\newcommand{\ucar}[1]{\underset{#1}{\times}}

\newcommand{\owed}[1]{\overset{#1}{\wedge}}

\newcommand{\balp}{\boldsymbol{\alp}}

\newcommand{\bmu}{\boldsymbol{\mu}}

\newcommand{\bsig}{\boldsymbol{\sig}}

\newcommand{\cA}{\mathcal{A}}

\newcommand{\cC}{\mathcal{C}}

\newcommand{\cE}{\mathcal{E}}

\newcommand{\cJ}{\mathcal{J}}

\newcommand{\cL}{\mathcal{L}}

\newcommand{\cT}{\mathcal{T}}

\newcommand{\by}{\boldsymbol{y}}

\newcommand{\bF}{\boldsymbol{F}}
\newcommand{\bG}{\boldsymbol{G}}

\newcommand{\bP}{\boldsymbol{P}}

\newcommand{\bU}{\boldsymbol{U}}

\newcommand{\bW}{\boldsymbol{W}}
\newcommand{\bX}{\boldsymbol{X}}
\newcommand{\bY}{\boldsymbol{Y}}

\newcommand{\sub}{\subset}

\newcommand{\wed}{\wedge}
\newcommand{\com}{\!\circ\!}
\newcommand{\ten}{\!\otimes\!}

\newcommand{\alp}{\alpha}
\newcommand{\bet}{\beta}
\newcommand{\gam}{\gamma}
\newcommand{\del}{\delta}
\newcommand{\eps}{\epsilon}
\newcommand{\zet}{\zeta}

\newcommand{\lam}{\lambda}
\newcommand{\sig}{\sigma}

\newcommand{\ome}{\omega}

\newcommand{\Lam}{\Lambda}

\newcommand{\vartht}{\vartheta}

\newcommand{\For}{{\Lambda}}
\newcommand{\Con}{{\mathcal{C}}}
\newcommand{\Hor}{{\mathcal{H}}}
\newcommand{\Var}{{\mathcal{V}}}
\newcommand{\Thd}{{\Theta}}

\title{\large{{\bf Gauge-natural field theories and Noether Theorems: canonical covariant
conserved currents}\thanks{
This paper is in final form and will not be submitted elsewhere.}}}
\author{{\normalsize  M. Palese and E. Winterroth\thanks{Both authors were partially supported by GNFM of INdAM. The second author was also
supported by University of Torino - {\em Dottorato di Ricerca in Matematica XVIII Ciclo}.}}
\\{\footnotesize Department of Mathematics,
University of Torino}
\\{\footnotesize via C. Alberto 10, 10123 Torino, Italy}\\ 
{\footnotesize e--mails: 
{\sc marcella.palese@unito.it, ekkehart@dm.unito.it}}}
\date{}
\begin{document}

\maketitle

\begin{abstract}

Recently we found that canonical gauge-natural superpotentials are obtained as global sections of the {\em reduced}
$(n-2)$-degree and
$(2s-1)$-order quotient sheaf on the fibered manifold $\bY_{\zet} \times_{\bX} 
\mathfrak{K}$, where $\mathfrak{K}$ is an appropriate subbundle of the vector bundle of (prolongations of)
infinitesimal right-invariant automorphisms $\bar{\Xi}$.  In this paper, we provide an alternative proof of the fact
that the naturality property
$\cL_{j_{s}\bar{\Xi}_{H}}\omega (\lambda ,
\mathfrak{K})=0$ holds true for the {\em new} Lagrangian $\omega (\lambda,
\mathfrak{K})$ obtained contracting the Euler--Lagrange form of the original Lagrangian with
$\bar{\Xi}_{V}\in
\mathfrak{K}$. We use as fundamental tools an invariant decomposition formula of vertical morphisms due to Kol\'a\v{r}
and the theory of iterated Lie derivatives of sections of fibered bundles. As a consequence, we  recover the existence
of a canonical generalized energy--momentum conserved tensor density associated with $\omega (\lambda ,
\mathfrak{K})$. 

\medskip


\end{abstract}

%-------------------------------------------------------------------------%
\section{Introduction}
%-------------------------------------------------------------------------%

Our general framework is the calculus of variations on finite order jets of 
{\em gauge-natural bundles} (\ie jet prolongations of fiber bundles associated 
to some gauge-natural prolongation of a principal bundle $\bP$ \cite{Ec81,KMS93}). 
Such geometric structures have been widely recognized to suitably describe 
so-called gauge-natural field theories, \ie physical theories in which
right-invariant infinitesimal automorphisms of the structure bundle $\bP$  
uniquely define the transformation laws of the fields themselves (see {\em e.g.} 
\cite{Ec81,FFP01,Ja03,KMS93} and references quoted therein).
In particular, we shall work within the differential setting of {\em finite order variational sequences
on gauge-natural bundles}.
In fact, it become  evident that the passage from
Lagrangians to Euler--Lagrange equations can be seen as a differential of a
complex (see \eg \cite{Tak79,Tul77,Vin84,Kru90}): the theory of finite order variational sequences provides 
then a suitable geometric framework for 
the Calculus of Variations. In this theory the Euler--Lagrange operator
is a differential morphism in a sequence of sheaves 
of vector spaces. Geometric objects like Lagrangians, momenta, 
Poincar\'e--Cartan forms, Helmholtz conditions, Jacobi equations, find a nice 
interpretation in the quotient spaces of the sequence of a given order. 

In the beginning of the second half of the past Century, to conveniently derive conserved
quantities for covariant field theories, it appeared necessary to define in a functorial and unique way
the lift of infinitesimal transformations of the basis manifolds to the bundle
of fields (namely bundles of tensor fields or tensor densities as
suitable representations of the action of infinitesimal space-time transformations on frame
bundles of a given order \cite{PaTe77}) \cite{AnBe51,Ber49,Ber58}. 
Such theories were also
called geometric or {\em natural} \cite{Tra67}. An important generalization of natural theories to gauge 
fields theories passed
through 
the concept of jet prolongation of a principal bundle and the
introduction of a very important geometric construction, namely the {\em gauge-natural bundle
functor} \cite{Ec81,KMS93}. 

In particular, P.G. Bergmann in \cite{Ber49} introduced what he called {\em generalized Bianchi identities} for geometric field
theories to get (after an integration by parts procedure) a consistent
equation involving local divergences within the first variation formula. It is well known that, following the Noether theory \cite{Noe18}, in
the classical Lagrangian formulation of field theories
the  description of symmetries and conserved quantities
consists in deriving from the invariance of the Lagrangian the existence of suitable 
conserved currents; in most relevant physical theories this currents are found to be the 
divergence of skew--symmetric
(tensor) densities, which are called {\em superpotentials} for the 
conserved currents themselves. It
is also well known that the importance of superpotentials relies on 
the fact that they can
be integrated to provide conserved quantities associated with the 
conserved currents {\em
via} the Stokes Theorem (see \eg \cite{FFP01,Mat03} and references therein).
Generalized Bergmann--Bianchi identities are in fact necessary 
and locally sufficient conditions for a Noether 
conserved current to be not only closed but also the divergence of a 
a superpotential 
along solutions of the  Euler--Lagrange equations. However, the problem of the general
covariance of such identities exists and it was already posed and partially investigated by
Anderson and Bergmann in
\cite{AnBe51}, where the invariance with respect to time coordinate transformations was
studied. This problem reflects obviously on the covariance of conserved quantities (see Remark
\ref{arbitrary1} below). Here we propose a way to deal with such open problems concerning
globality aspects. For the relevance of the latter ones also in quantum field theories, see
\eg the preprints \cite{BGMS04-05}.

In \cite{FPV98a} 
a representation of symmetries in finite order variational sequences was 
provided by means of the introduction of the {\em
variational Lie derivative}, \ie the induced quotient operator acting 
on equivalence classes of forms in the variational
sequence. In \cite{FFP01} the theory of Noether conserved currents 
and superpotentials was tackled by using such
representations for natural and gauge-natural Lagrangian field theories. 
Recently, further developments have been achieved
concerning a canonical covariant derivation of Noether conserved quantities 
and global superpotentials \cite{PaWi03,PaWi04}. On the other hand the {\em second variation} 
of the action functional can be conveniently represented in the 
finite order variational sequence framework in terms of iterated variational 
Lie derivatives of the Lagrangian  with respect to 
vertical parts of gauge-natural lifts of principal infinitesimal 
automorphisms. In particular, in \cite{FrPa01,FPV02} the second variational 
derivative has been represented and
related with the {\em generalized Jacobi morphism}. Furthermore, the gauge-natural structure of 
the theories under consideration enables us to
define the {\em generalized gauge-natural Jacobi morphism} where the {\em variation vector fields} 
are Lie derivatives 
of sections of the gauge-natural bundle with respect to gauge-natural lifts. 

In this paper we use representations of the Noether Theorems given in \cite{FPV98a}; in particular we specialize in a new way  the Second Noether Theorem for gauge-natural theories  by means of the Jacobi morphism \cite{PaWi03,PaWi04} and show that the Second Noether Theorem plays a
fundamental role in the derivation of canonical covariant conserved quantities in gauge-natural
field theories (see Remark \ref{application} below). In fact, the {\em indeterminacy} 
appearing in the derivation of gauge-natural conserved charges (for a review, see 
the interesting papers \cite{GoMa03,Mat03}) - \ie the difficulty of 
relating in a natural way infinitesimal gauge transfomations with 
infinitesimal transformations of the basis manifold - can be solved by requiring the second 
variational derivative to vanish {\em as well}\, \cite{PaWi04}. 
Moreover, for gauge-natural field theories, here we stress that generalized Bergmann--Bianchi identities hold true in a canonical covariant
way {\em if and only if} the second variational derivative - with respect to vertical parts of gauge-natural lifts - of the
Lagrangian vanishes \cite{PaWi03}.  As a quite strong consequence, for any gauge-natural invariant field theory we find that the above mentioned indeterminacy can be {\em always} solved canonically.

As a consequence of the Second Noether Theorem, we further show that 
there exists a covariantly conserved current associated with the Lagrangian obtained by contracting the
Euler--Lagrange morphism with a gauge-natural Jacobi vector field.

%---------------------------------------------------------------------------------------------------
\section{Finite order jets of gauge-natural bundles}
%---------------------------------------------------------------------------------------------------

We recall some basic facts about jet spaces \cite{KMS93,Sau89}.
Our framework is a fibered manifold $\pi : \bY \to \bX$,
with $\dim \bX = n$ and $\dim \bY = n+m$.

For $s \geq q \geq 0$ integers we are concerned with the $s$--jet space $J_s\bY$ of 
$s$--jet prolongations of (local) sections
of $\pi$; in particular, we set $J_0\bY \equiv \bY$. We recall the natural fiberings
$\pi^s_q: J_s\bY \to J_q\bY$, $s \geq q$, $\pi^s: J_s\bY \to \bX$, and,
among these, the {\em affine\/} fiberings $\pi^{s}_{s-1}$.
We denote by $V\bY$ the vector subbundle of the tangent
bundle $T\bY$ of vectors on $\bY$ which are vertical with respect
to the fibering $\pi$.

Greek
indices $\sig ,\mu ,\dots$ run from $1$ to $n$ and they label basis
coordinates, while
Latin indices $i,j,\dots$ run from $1$ to $m$ and label fibre coordinates,
unless otherwise specified. 
We denote multi--indices of dimension $n$ by boldface Greek letters such as
$\balp = (\alp_1, \dots, \alp_n)$, with $0 \leq \alp_\mu$,
$\mu=1,\ldots,n$; by an abuse
of notation, we denote by $\sig$ the multi--index such that
$\alp_{\mu}=0$, if $\mu\neq \sig$, $\alp_{\mu}= 1$, if
$\mu=\sig$.
We also set $|\balp| \byd \alp_{1} + \dots + \alp_{n}$ and $\balp ! \byd
\alp_{1}! \dots \alp_{n}!$.
The charts induced on $J_s\bY$ are denoted by $(x^\sig,y^i_{\balp})$, with $0
\leq |\balp| \leq s$; in particular, we set $y^i_{\bf{0}}
\equiv y^i$. The local vector fields and forms of $J_s\bY$ induced by
the above coordinates are denoted by $(\der^{\balp}_i)$ and $(d^i_{\balp})$,
respectively.

For $s\geq 1$, we consider the natural complementary fibered
morphisms over $J_s\bY \to J_{s-1}\bY$ (see \eg \cite{Kru90,Kru93,Vit98}):
\beq
\mathcal{D} : J_s\bY \ucar{\bX} T\bX \to TJ_{s-1}\bY \,,
\qquad
\vartht : J_{s}\bY \ucar{J_{s-1}\bY} TJ_{s-1}\bY \to VJ_{s-1}\bY \,,
\eeq
with coordinate expressions, for $0 \leq |\balp| \leq s-1$, given by
\beq
\mathcal{D} &= d^\lam\ten {\mathcal{D}}_\lam = d^\lam\ten
(\der_\lam + y^j_{\balp+\lam}\der_j^{\balp}) \,,
\vartht &= \vartht^j_{\balp}\ten\der_j^{\balp} =
(d^j_{\balp}-y^j_{{\balp}+\lam}d^\lam)
\ten\der_j^{\balp} \,.
\eeq

The morphisms above induce the following natural splitting (and its dual):
\bEq
\label{jet connection}
J_{s}\bY\ucar{J_{s-1}\bY}T^*J_{s-1}\bY =\left(
J_s\bY\ucar{J_{s-1}\bY}T^*\bX\right) \oplus\cC^{*}_{s-1}[\bY]\,,
\eEq
where $\cC^{*}_{s-1}[\bY] \byd \im \vartht_s^*$ and
$\vartht_s^* : J_s\bY \ucar{J_{s-1}\bY} V^*J_{s-1}\bY \to
J_s\bY \ucar{J_{s-1}\bY} T^*J_{s-1}\bY \,$.

If $f: J_{s}\bY \to \R$ is a function, then we set
$D_{\sig}f$ $\byd \mathcal{D}_{\sig} f$,
$D_{\balp+\sig}f$ $\byd D_{\sig} D_{\balp}f$, where $D_{\sig}$ is
the standard {\em formal derivative}.
Given a vector field $\Xi : J_{s}\bY \to TJ_{s}\bY$, the splitting
\eqref{jet connection} yields $\Xi \, \com \, \pi^{s+1}_{s} = \Xi_{H} + \Xi_{V}$
where, if $\Xi = \Xi^{\gam}\der_{\gam} + \Xi^i_{\balp}\der^{\balp}_i$, then we
have $\Xi_{H} = \Xi^{\gam}D_{\gam}$ and
$\Xi_{V} = (\Xi^i_{\balp} - y^i_{\balp + \gam}\Xi^{\gam}) 
\der^{\balp}_{i}$. We shall call $\Xi_{H}$ and $\Xi_{V}$ the 
horizontal and the vertical part of $\Xi$, respectively.

The splitting
\eqref{jet connection} induces also a decomposition of the
exterior differential on $\bY$,
$(\pi^{s}_{s-1})^*\com \,d = d_H + d_V$, where $d_H$ and $d_V$
are defined to be the {\em horizontal\/} and {\em vertical differential\/}.
The action of $d_H$ and $d_V$ on functions and $1$--forms
on $J_s\bY$ uniquely characterizes $d_H$ and $d_V$ (see, {\em e.g.},
\cite{Sau89,Vit98} for more details).
A {\em projectable vector field\/} on $\bY$ is defined to be a pair
$(u,\xi)$, where $u:\bY \to T\bY$ and $\xi: \bX \to T\bX$
are vector fields and $u$ is a fibered morphism over $\xi$.
If there is no danger of confusion, we will denote simply by $u$ a
projectable vector field $(u,\xi)$.
A projectable vector field $(u,\xi)$
can be conveniently prolonged to a projectable vector field
$(j_{s}u, \xi)$; coordinate expression can be found \eg in 
\cite{Kru90,Sau89,Vit98}.

%---------------------------------------------------------------
\subsection{Gauge-natural bundles}
%---------------------------------------------------------------

Let $\bP\to\bX$ be a principal bundle with structure group $\bG$.
Let $r\leq k$ be integers and $\bW^{(r,k)}\bP$ $\byd$ $J_{r}\bP\ucar{\bX}L_{k}(\bX)$, 
where $L_{k}(\bX)$ is the bundle of $k$--frames 
in $\bX$ \cite{Ec81,KMS93}, $\bW^{(r,k)}\bG \byd J_{r}\bG\odot GL_{k}(n)$
the semidirect product with respect to the action of $GL_{k}(n)$ 
on $J_{r}\bG$ given by the 
jet composition and $GL_{k}(n)$ is the group of $k$--frames 
in $\R^{n}$. Here we denote by $J_{r}\bG$ the space of $(r,n)$-velocities on $\bG$ \cite{KMS93}.
The bundle $\bW^{(r,k)}\bP$ is a principal bundle over $\bX$ with structure group
$\bW^{(r,k)}\bG$.
Let $\bF$ be any manifold and $\zet:\bW^{(r,k)}\bG\ucar{}\bF\to\bF$ be 
a left action of $\bW^{(r,k)}\bG$ on $\bF$. There is a naturally defined 
right action of $\bW^{(r,k)}\bG$ on $\bW^{(r,k)}\bP \times \bF$ so that
 we can associate in a standard way
to $\bW^{(r,k)}\bP$ the bundle, on the given basis $\bX$,
$\bY_{\zet} \byd \bW^{(r,k)}\bP\times_{\zet}\bF$.

\bDf
We say $(\bY_{\zet},\bX,\pi_{\zet};\bF,\bG)$ to be the 
{\em gauge-natural bundle} of order 
$(r,k)$ associated to the principal bundle $\bW^{(r,k)}\bP$ 
by means of the left action $\zet$ of the group 
$\bW^{(r,k)}\bG$ on the manifold $\bF$ \cite{Ec81,KMS93}. 
\END\eDf

\bRm
A principal automorphism $\Phi$ of $\bW^{(r,k)}\bP$ induces an 
automorphism of the gauge-natural bundle by:
\bEq
\Phi_{\zet}:\bY_{\zet}\to\bY_{\zet}: [(j^{x}_{r}\gam,j^{0}_{k}t), 
\hat{f}]_{\zet}\mto [\Phi(j^{x}_{r}\gam,j^{0}_{k}t), 
\hat{f}]_{\zet}\,, 
\eEq
where $\hat{f}\in \bF$ and $[\cdot, \cdot]_{\zet}$ is the equivalence class
induced by the action $\zet$.
\END\eRm
\bDf
We define the {\em vector}
bundle over $\bX$ of right--invariant infinitesimal automorphisms of $\bP$
by setting $\cA = T\bP/\bG$. 

We also define the {\em vector} bundle  over $\bX$ of right invariant 
infinitesimal automorphisms of $\bW^{(r,k)}\bP$ by setting 
$\cA^{(r,k)} \byd T\bW^{(r,k)}\bP/\bW^{(r,k)}\bG$ ($r\leq k$).
\END\eDf

Denote by $\cT_{\bX}$ and $\cA^{(r,k)}$ the sheaf of
vector fields on $\bX$ and the sheaf of right invariant vector fields 
on $\bW^{(r,k)}\bP$, respectively. A functorial mapping $\mathfrak{G}$ is defined 
which lifts any right--invariant local automorphism $(\Phi,\phi)$ of the 
principal bundle $W^{(r,k)}\bP$ into a unique local automorphism 
$(\Phi_{\zet},\phi)$ of the associated bundle $\bY_{\zet}$. 
Its infinitesimal version associates to any $\bar{\Xi} \in \cA^{(r,k)}$,
projectable over $\xi \in \cT_{\bX}$, a unique {\em projectable} vector field 
$\hat{\Xi} \byd \mathfrak{G}(\bar{\Xi})$ on $\bY_{\zet}$, the {\em gauge-natural lift}, in the 
following way:
\bEq
\mathfrak{G} : \bY_{\zet} \ucar{\bX} \cA^{(r,k)} \to T\bY_{\zet} \,:
(\by,\bar{\Xi}) \mto \hat{\Xi} (\by) \,,
\eEq
where, for any $\by \in \bY_{\zet}$, one sets: $\hat{\Xi}(\by)=
\frac{d}{dt} [(\Phi_{\zet \,t})(\by)]_{t=0}$,
and $\Phi_{\zet \,t}$ denotes the (local) flow corresponding to the 
gauge-natural lift of $\Phi_{t}$.

This mapping fulfils the following properties (see \cite{KMS93}):
\begin{enumerate}
\item $\mathfrak{G}$ is linear over $id_{\bY_{\zet}}$;
\item we have $T\pi_{\zet}\circ\mathfrak{G} = id_{T\bX}\circ 
\bar{\pi}^{(r,k)}$, 
where $\bar{\pi}^{(r,k)}$ is the natural projection
$\bY_{\zet}\ucar{\bX} 
\cA^{(r,k)} \to T\bX$;
\item for any pair $(\bar{\Lam},\bar{\Xi})$ $\in$
$\cA^{(r,k)}$, we have
$\mathfrak{G}([\bar{\Lam},\bar{\Xi}]) = [\mathfrak{G}(\bar{\Lam}), \mathfrak{G}(\bar{\Xi})]$.
\end{enumerate}

%----------------------------------------------------------------------%
\subsection{Lie derivative of sections}\label{2.2}
%----------------------------------------------------------------------%

\bDf
Let $\gam$ be a (local) section of $\bY_{\zet}$, $\bar{\Xi}$ 
$\in \cA^{(r,k)}$ and $\hat\Xi$ its gauge-natural lift. 
Following \cite{KMS93} we
define the {\em 
generalized Lie derivative} of $\gam$ along the vector field 
$\hat{\Xi}$ to be the (local) section $\pounds_{\bar{\Xi}} \gam : \bX \to V\bY_{\zet}$, 
given by
$\pounds_{\bar{\Xi}} \gam = T\gam \circ \xi - \hat{\Xi} \circ \gam$.\END
\eDf

\bRm\label{lie}
The Lie derivative operator acting on sections of gauge-natural 
bundles satisfies the following 
properties:
\begin{enumerate}\label{lie properties}
\item for any vector field $\bar{\Xi} \in \cA^{(r,k)}$, the 
mapping $\gam \mto \pounds_{\bar{\Xi}}\gam$ 
is a first--order quasilinear differential operator;
\item for any local section $\gam$ of $\bY_{\zet}$, the mapping 
$\bar{\Xi} \mto \pounds_{\bar{\Xi}}\gam$ 
is a linear differential operator;
\item we can regard $\pounds_{\bar{\Xi}}: J_{1}\bY_{\zet} \to V\bY_{\zet}$ 
as a morphism over the
basis $\bX$. By using the canonical 
isomorphisms $VJ_{s}\bY_{\zet}\simeq J_{s}V\bY_{\zet}$ for all $s$, we have
$\pounds_{\bar{\Xi}}[j_{s}\gam] = j_{s} [\pounds_{\bar{\Xi}} \gam]$,
for any (local) section $\gam$ of $\bY_{\zet}$ and for any (local) 
vector field $\bar{\Xi}\in \cA^{(r,k)}$. Furthermore, for gauge-natural lifts, the fundamental relation hold true:
\bEq\label{ksivu}
\hat{\Xi}_V\byd\mathfrak{G}(\bar{\Xi})_V=- \pounds_{\bar{\Xi}}\,.
\eEq
\end{enumerate}\END
\eRm

%----------------------------------------------------------------------------
\section{Variational sequences and Noether Theorems} 
%------------------------------------------------------------------------------

For the sake of simplifying notation, sometimes, we will omit the subscript $\zet$, so 
that all our considerations shall refer to $\bY$ as a gauge-natural 
bundle as defined above.

For convenience of the reader, we sketch the connection of the purely differential setting 
of variational sequences with
the classical integral presentation of Calculus of Variations, although the two
approaches (differential and integral one) are completely independent,
even if the latter provided the motivation to the former from an
historical viewpoint.

In the formulation of variational problems on jet spaces of a fibered manifold $\bY\to\bX$, 
with $n=\textstyle{dim} \bX$ and $m=\textstyle{dim} \bY - n$ 
(see \eg \cite{GoSt73,Kol83,Pal68,Sau89}), it is well known that, 
given an $s$-th order Lagrangian $\lambda\in\Hor^n_s$, the {\em action} of $\lambda$ along a section 
$\gam: \bU\to\bY$, on an oriented open subset 
$\bU$ of $\bX$ with compact closure and regular boundary, is defined to be the real number
\beq
\int_{\bU}(j_s \gam)^*\lambda\, .
\eeq
A {\em variation vector field} is a vertical vector field $u \colon
\bY\to V\bY$ defined along $\gam(\bU)$.
A local section $\gam \colon\bU\to\bY$ is said to be {\em critical} if, for each
variation vector field with flow $\phi_t$, we have
$$\delta\int_{\bU}(j_s \phi_t\circ j_s \gam)^*\lambda=0\,,$$ where $\delta$ is
the Fr\'{e}chet derivative with respect to the parameter $t$, at $t=0$. It is
easy to see that the previous integral expression is equal to
$\int_{\bU}(j_s \gam)^*\text{L}_{j_s u}\lambda=0$ for each variation vector field
$u$, where $\text{L}_{j_s u}$ is the Lie derivative operator. For each variation vector field $u$ satisfying suitable
boundary conditions, since 
$L_{j_s u}\lambda=i_{j_s u}d\lambda$, as an application of the Stokes Theorem, we find
that the above equation is equivalent to
$\int_{\bU}(j_{2s}\gam)^*(i_u E_{d\lambda})=0$, where $E_{d\lambda}$ is the {\em generalized Euler--Lagrange operator} 
associated with $\lam$ (see later).  
Finally, by virtue of the
fundamental Lemma of the Calculus of Variations the above condition is
equivalent to $E_{d\lambda}\circ j_{2s}\gam =0$, known as the Euler--Lagrange equations (see \eg the review in
\cite{Kru90}).

\medskip

Let us now construct the Krupka's finite order variational sequence. 

According to
\cite{Kru90,Vit98}, the fibered splitting
\eqref{jet connection} yields the {\em sheaf splitting}
$\Hor^{p}_{(s+1,s)}$ $=$ $\bigoplus_{t=0}^p$
$\Con^{p-t}_{(s+1,s)}$ $\wed\Hor^{t}_{s+1}$, which restricts to the inclusion
$\For^{p}_s$ $\sub$ $\bigoplus_{t=0}^{p}$
$\Con^{p-t}{_s}\wed\Hor^{t,}{_{s+1}^{h}}$,
where $\Hor^{p,}{_{s+1}^{h}}$ $\byd$ $h(\For^{p}_s)$ for $0 < p\leq 
n$ and the surjective map
$h$ is defined to be the restriction to $\For^{p}_{s}$ of the projection of
the above splitting onto the non--trivial summand with the highest
value of $t$.
By an abuse of notation, let us denote by $d\ker h$ the sheaf
generated by the presheaf $d\ker h$ in the standard way.
We set $\Thd^{*}_{s}$ $\byd$ $\ker h$ $+$
$d\ker h$.

In \cite{Kru90} 
it was proved that the following {\em $s$--th order
variational sequence} associated with the fibered manifold
$\bY\to\bX$ is an exact resolution of the constant sheaf $\R_{\bY}$ over $\bY$:
\beq
\diagramstyle[size=1.3em]
\begin{diagram}
0 & \rTo & \R_{\bY} & \rTo & \For^{0}_s & \rTo^{\cE_{0}} &
\For^{1}_s/\Thd^{1}_s & \rTo^{\cE_{1}} & \For^{2}_s/\Thd^{2}_s & \rTo^{\cE_{2}} &
\dots & \rTo^{\cE_{I-1}} & \For^{I}_s/\Thd^{I}_s & \rTo^{\cE_{I}} &
\For^{I+1}_s & \rTo^{d} & 0 \, ,
\end{diagram}
\eeq
where the integer $I$ depends on the dimension of the fibers of $\bY$ (see \cite{Kru90}).

For practical purposes we 
shall limit ourselves to consider the truncated variational sequence introduced by Vitolo in
\cite{Vit98}: 
\beq
\diagramstyle[size=1.3em]
\begin{diagram}
0 &\rTo & \R_{Y} &\rTo & \Var^{0}_s & \rTo^{\cE_0} &
\Var^{1}_{s} & \rTo^{\cE_{1}} & \dots  & \rTo^{\cE_{n}} &
\Var^{n+1}_{s}  & \rTo^{\cE_{n+1}} & \cE_{n+1}(\Var^{n+1}_{s})  
& \rTo^{\cE_{n+2}} & 
0 \,,
\end{diagram}
\eeq
where, following \cite{Vit98}, the sheaves $\Var^{p}_{s}\byd 
\Con^{p-n}_{s}\wed\Hor^{n,}{_{s+1}^h}/h(d\ker h)$ with $0\leq p\leq n+2$ are 
suitable representations of the corresponding quotient 
sheaves in the variational sequence by means of sheaves of sections of tensor
bundles. 

Let $\alp\in\Con^{1}_s\wed\Hor^{n,}{_{s+1}^h} 
\sub \Var^{n+1}_{s+1}$. Then there is a unique pair of
sheaf morphisms (\cite{Kol83,KoVi03,Vit98})
\bEq\label{first variation}
E_{\alp} \in \Con^{1}_{(2s,0)}\wed\Hor^{n,}{_{2s+1}^{h}} \,,
\qquad
F_{\alp} \in \Con^{1}_{(2s,s)} \wed \Hor^{n,}{_{2s+1}^h} \,,
\eEq
such that 
$(\pi^{2s+1}_{s+1})^*\alp=E_{\alp}-F_{\alp}$
and $F_\alp$ is {\em locally} of the form $F_{\alp} = d_{H}p_{\alp}$, with $p_{\alp}
\in \Con^{1}_{(2s-1,s-1)}\wed\Hor^{n-1}{_{2s}}$.

We shall now introduce a - for our purposes - fundamental morphism, denoted by $K_{\eta}$, represented by
Vitolo in
\cite{Vit98} and further studied by Kol\'a\v{r} and Vitolo in \cite{KoVi03}. 

Let then
$\eta\in\Con^{1}_{s}\wed\Con^{1}_{(s,0)}\wed\Hor^{n,}{_{s+1}^{h}}\sub 
\Var^{n+2}_{s+1}$;
then there is a unique morphism
$$
K_{\eta} \in \Con^{1}_{(2s,s)}\otimes\Con^{1}_{(2s,0)}\wed\Hor^{n,}{_{2s+1}^{h}}
$$
such that, for all $\Xi:\bY\to V\bY$,
$E_{{j_{s}\Xi}\rfloor \eta} = C^{1}_{1} (j_{2s}\Xi\ten K_{\eta})$,
where $C^1_1$ stands for tensor
contraction on the first factor and $\rfloor$ denotes inner product (see \cite{KoVi03,Vit98}). 
Furthermore, there is a unique pair of sheaf morphisms
\bEq\label{second}
H_{\eta} \in 
\Con^{1}_{(2s,s)}\wed\Con^{1}_{(2s,0)}\wed\Hor^{n,}{_{2s+1}^{h}} \,,
\quad
G_{\eta} \in \Con^{2}_{(2s,s)}\wed\Hor^{n,}{_{2s+1}^{h}} \,,
\eEq
such that 
${(\pi^{2s+1}_{s+1})}^*\eta = H_{\eta} - G_{\eta}$ and $H_{\eta} 
= \frac{1}{2} \, A(K_{\eta})$,
where $A$ stands for antisymmetrisation.
Moreover, $G_{\eta}$ is {\em locally} of the type $G_{\eta} = d_H q_{\eta}$, 
where 
$q_{\eta} \in \Con^{2}_{(2s-1,s-1)}\wed\Hor^{n-1}{_{2s}}$; hence 
$[\eta]=[H_{\eta}]$ \cite{KoVi03,Vit98}. 

\bRm
A section $\lam\in\Var^{n}_s$ is just a Lagrangian of order 
$(s+1)$ of 
the standard literature. 
Furthermore
$\cE_{n}(\lam) \in \Var^{n+1}_{s}$ coincides with the standard higher 
order Euler--Lagrange morphism associated with $\lam$. Let $\gam \in \For^{n+1}_{s}$.
The morphism $H_{hd\gam}\equiv H_{[\cE_{n+1}(\gam)]}$, where square 
brackets denote equivalence class, is called the {\em
generalized Helmholtz morphism}; its kernel coincides with Helmholtz conditions of local
variationality. We shall integrate by
parts the morphism $K_{\eta}$ to provide a suitable representation of the {\em generalized
Jacobi morphism} associated with $\lam$ \cite{FrPa01,FPV02,PaWi03,PaWi04}.
\END
\eRm

The standard Lie derivative of fibered morphisms with respect to a 
projectable vector field $j_{s}\Xi$ passes to the quotient in the 
variational sequence, so defining a new quotient operator 
(introduced in \cite{FPV98a}), the {\em variational Lie derivative} 
$\cL_{j_{s}\Xi}$, acting on equivalence classes of fibered morphisms 
which are sections of the quotient sheaves in the variational 
sequence. Thus variational Lie derivatives of generalized Lagrangians 
or Euler--Lagrange morphisms can be conveniently represented as 
equivalence classes in $\Var^{n}_{s}$ and $\Var^{n+1}_{s}$. In 
particular, the following two results hold true \cite{FPV98a}, to which for evident 
reasons we will refer as the First and the Second Noether Theorem, respectively.

\bTh\label{noether I}
Let $[\alp] = h(\alp)$ $\in$ $\Var^{n}_{s}$. Then we
have {\em locally} (up to pull-backs)
\beq
\cL_{j_{s}\Xi}(h(\alp)) =
\Xi_{V} \rfloor \cE_{n}(h(\alp))+
d_{H}(j_{2s}\Xi_{V} \rfloor p_{d_{V}h(\alp)}+ \xi \rfloor h(\alp))\,.
\eeq
\eTh

\bTh\label{GeneralJacobi}
Let $\alp\in\For^{n+1}_{s}$. Then we have {\em globally} (up to pull-backs)
\beq 
\cL_{j_{s}\Xi} [\alp] =
\cE_{n}({j_{s+1}\Xi_{V} \rfloor h(\alp)}) +
C^{1}_{1}(j_{s}\Xi_{V}\ten K_{hd\alp}) \,.
\eeq
\eTh

Notice that the Second Noether Theorem as formulated above, is represented in terms of the
morphism $K_{hd\alp}$.

%-----------------------------------------------------------------------------%
\subsection{Noether conserved currents}
%-----------------------------------------------------------------------------%

In the following we assume that the field equations are generated by
means of a variational principle from a Lagrangian which is
gauge-natural invariant, \ie invariant with respect to any
gauge-natural lift of infinitesimal right invariant vector fields.
Both the Noether Theorems take a quite particular form in the case of gauge-natural 
Lagrangian field theories (see \eg  \cite{FFP01,Mat03}) due to the 
fact that the generalized Lie derivative of sections of the 
gauge-natural bundles has specific linearity properties recalled in Subsection \ref{2.2} and it
is  related with the vertical part of gauge-natural lifts  by Eq. \eqref{ksivu}. 

\bDf\label{gn}
Let $(\hat{\Xi},\xi)$ be a projectable vector field on $\bY_{\zet}$.
Let $\lam \in \Var^{n}_{s}$
be a generalized Lagrangian. We say $\hat{\Xi}$ to be a {\em symmetry\/}
of $\lam$ if $\cL_{j_{s+1}\hat{\Xi}}\,\lam = 0$.

We say $\lam$ to be a
{\em gauge-natural invariant Lagrangian} if the gauge-natural lift
$(\hat{\Xi},\xi)$ of {\em any} vector
field $\bar{\Xi} \in \cA^{(r,k)}$ is a  symmetry for
$\lam$, \ie if $\cL_{j_{s+1}\bar{\Xi}}\,\lam = 0$.
In this case the projectable vector field
$\hat{\Xi}\equiv \mathfrak{G}(\bar{\Xi})$ is
called a {\em gauge-natural symmetry} of $\lam$.\END
\eDf

In the following we rephrase the First Noether Theorem in the case of gauge-natural
Lagrangians.

\bPr
\label{symmetry of L}
Let $\lam \in \Var^{n}_{s}$ be a gauge-natural Lagrangian and
$(\hat{\Xi},\xi)$
a gauge-natural symmetry of $\lam$. Then we have
$
0 = - \pounds_{\bar{\Xi}} \rfloor \cE_{n}(\lam)
+d_{H}(-j_{s}\pounds_{\bar{\Xi}}
\rfloor p_{d_{V}\lam}+ \xi \rfloor \lam) $.
Suppose that
$(j_{2s+1}\sig)^{*}(- \pounds_{\bar{\Xi}} \rfloor \cE_{n}(\lam)) = 0$.
Then, the $(n-1)$--form
$\eps = - j_{s}\pounds_{\bar{\Xi}} \rfloor p_{d_{V}\lam}+ \xi \rfloor \lam$
fulfills the equation $d ((j_{2s}\sig)^{*}(\eps)) = 0$.
\ePr

If $\sig$ is a critical section for $\cE_{n}(\lam)$, \ie
$(j_{2s+1}\sig)^{*}\cE_{n}(\lam) = 0$, the above equation
admits a physical interpretation as a so-called {\em weak conservation law}
for the density associated with $\eps$ and the associated sheaf morphism $\eps: 
J_{2s}\bY_{\zet} \ucar{\bX} VJ_{2s}\cA^{(r,k)} \to 
\cC_{2s}^{*}[\cA^{(r,k)}]\ten\cC_{0}^{*}[\cA^{(r,k)}]
\wed (\owed{n-1} T^{*}\bX)$
is said to be a {\em gauge-natural weakly conserved current}.

\bRm\label{arbitrary1}
We stress that such a Noether conserved current {\em is not} uniquely defined, {\em even up to
divergences}. In fact, it depends on the choice of $p_{d_{V}\lam}$, which in  general is not
unique - even up to divergences - {\em depending on the fixing of suitable connections} used
to derive it in an invariant way (see \cite{Kol83,Vit98} and references quoted therein).
\END\eRm

%-----------------------------------------------------------------------------------
\section{Variations and generalized Jacobi morphisms}
%-----------------------------------------------------------------------------------

We consider {\em formal variations} of a morphism as  
{\em multiparameter deformations} and relate
 the second variational derivative of 
the Lagrangian $\lam$ to the
Lie derivative of the associated Euler--Lagrange morphism and in turn to the generalized
Bergmann--Bianchi morphism; see \cite{PaWi03} for details.

Let $\alp: J_{s}\bY\to
\owed{p}T^*J_{s}\bY$ and let $L_{j_{s}\Xi_{k}}$ be the Lie derivative 
operator acting on differential fibered morphism.
Let 
 $\Xi_{k}$, $1\leq k\leq i$, be (vertical) variation 
vector fields on $\bY$ in the sense of \cite{FrPa01,FPV02,PaWi03}. We define the
$i$--th formal variation of the morphism $\alp$ to be the operator:
$\del^{i} \alp = L_{j_{s}\Xi_{1}} \ldots L_{j_{s}\Xi_{i}} \alp$.

\bDf
Let $\alp\in (\Var^{n}_{s})_{\bY}$ and $\cL_{\Xi_{i}}$ the {\em variational Lie derivative} \cite{FPV98a} 
operator with respect to the variation vector field $\Xi_{i}$. 

\noindent We define the  {\em $i$--th variational
derivative} operator as follows: $\del^{i}[\alp]\byd [\del^{i}\alp]$  $=$ $[L_{\Xi_{i}} \ldots
L_{\Xi_{1}}\alp]$ $=$ $\cL_{\Xi_{i}} \ldots 
\cL_{\Xi_{1}}[\alp]$.\END
\eDf

It is clear that the first variational derivative is noting but the variational 
Lie derivative with respect to vertical parts of (gauge-natural lifts) of vector fields. Analogously, the second
variational derivative is nothing but the iterated (twice) variational Lie derivative; thus it can be expressed by
means of the Noether Theorems. As a straightforward consequence  the following
characterization of the second variational derivative of a generalized Lagrangian 
in the variational sequence holds true \cite{PaWi03}.

\bPr\label{x}
Let $\lam\in (\Var^{n}_{s})_{\bY}$ and let $\Xi$ be a variation vector 
field; then we have
\bEq
\del^{2}\lam = [\cE_{n}(j_{2s}\Xi \rfloor h\del\lam)
+C^{1}_{1} (j_{2s}\Xi \ten K_{hd\del\lam})] \,.
\eEq
\ePr

%------------------------------------------------------------------------------------------------
\subsection{Generalized {\em gauge-natural} Jacobi morphisms}
%------------------------------------------------------------------------------------------------

Let $\lam$ be a Lagrangian and $\bar{\Xi}$ a variation vector field. Let us set
$\chi(\lam,\mathfrak{G}(\bar{\Xi})_{V})$ $\byd$ 
 $C^{1}_{1} (j_{2s}\hat{\Xi}$ $\ten$ $K_{hd\cL_{j_{2s}\bar{\Xi}_V}\lam})$ $\equiv$  $E_{j_{s}\hat{\Xi}\rfloor
hd\cL_{j_{2s+1}\bar{\Xi}_V}\lam}$. Let 
$D_{H}$ be the horizontal differential on $\bY_{\zet}\ucar{\bX}V\cA^{(r,k)}$. Since $D_{H}
\chi(\lam,\mathfrak{G}(\bar{\Xi})_{V})$ $=$ $0$, by applying a global decomposition
formula for vertical morphisms due to Kol\'a\v{r} \cite{Kol83}, as a consequence of linearity properties
of both $\chi(\lam, \mathfrak{G}(\bar{\Xi})_{V})$ and 
 the Lie derivative operator $\pounds$,  
from Proposition \ref{x} we deduce what follows.

\bLm
We have:
\beq
(\pi^{4s+1}_{2s+1})^{*}\chi(\lam,\mathfrak{G}(\bar{\Xi})_{V}) = E_{\chi(\lam,\mathfrak{G}(\bar{\Xi})_{V})} +
F_{\chi(\lam,\mathfrak{G}(\bar{\Xi})_{V})}\,,
\eeq 
where
\beq
E_{\chi(\lam,\mathfrak{G}(\bar{\Xi})_{V}}: J_{4s}\bY_{\zet}\ucar{\bX}VJ_{4s}\cA^{(r,k)} \to 
\Con^{*}_{0}[\cA^{(r,k)}]\ten\Con^{*}_{0}[\cA^{(r,k)}]\wed (\owed{n}T^{*}\bX) \,,
\eeq
and locally, $F_{\chi(\lam,\mathfrak{G}(\bar{\Xi})_{V})} = 
D_{H}M_{\chi(\lam,\mathfrak{G}(\bar{\Xi})_{V})}$, with
\beq
M_{\chi(\lam,\mathfrak{G}(\bar{\Xi})_{V})}: J_{4s-1}\bY_{\zet}\ucar{\bX}VJ_{4s-1}\cA^{(r,k)}\to 
\Con^{*}_{2s-1}[\cA^{(r,k)}]\ten \Con^{*}_{0}[\cA^{(r,k)}]\wed
(\owed{n-1}T^{*}\bX)\,.
\eeq
\eLm

\bDf
We call the morphism $\cJ(\lam,\mathfrak{G}(\bar{\Xi})_{V})$ $\byd$
$E_{\chi(\lam,\mathfrak{G}(\bar{\Xi})_{V})}$  the {\em gauge-natural generalized Jacobi 
morphism} associated with the Lagrangian $\lam$ and the variation vector field
$\mathfrak{G}(\bar{\Xi})_{V}$.
\END\eDf

The morphism $\cJ(\lam,\mathfrak{G}(\bar{\Xi})_{V})$ is a 
{\em linear} morphism with respect to the projection 
$J_{4s}\bY_{\zet}\ucar{\bX}VJ_{4s}\cA^{(r,k)} \to J_{4s}\bY_{\zet}$.

As a consequence of Theorem \ref{GeneralJacobi} 
 and Proposition \ref{x} we have the
following characterization of the Second Noether Theorem for gauge-natural invariant
Lagrangian field theories in terms of the second variational derivative (see \cite{PaWi03} for the proof in detail).

\bTh\label{comparison}
Let $\del^{2}_{\mathfrak{G}}\lam$ be the variation of $\lam$ with respect to vertical parts
of gauge-natural lifts of infinitesimal principal automorphisms. We have:
\bEq
\cL_{\mathfrak{G}(\bar{\Xi})_V}\cL_{\mathfrak{G}(\bar{\Xi})_V}\byd \del^{2}_{\mathfrak{G}}\lam 
= \cJ(\lam,\mathfrak{G}(\bar{\Xi})_{V}) \,.
\eEq
Furthermore:
\bEq
\cJ(\lam,\mathfrak{G}(\bar{\Xi})_{V})=\mathfrak{G}(\bar{\Xi})_{V}\rfloor
\cE_{n}(\mathfrak{G}(\bar{\Xi})_{V}\rfloor\cE_{n}(\lam))
=
\cE_{n}(\mathfrak{G}(\bar{\Xi})_{V}\rfloor 
h(d\del\lam))\,. 
\eEq
\eTh

This result generalizes a classical result due to Goldschmidt and Sternberg
\cite{GoSt73} relating the Hessian with the Jacobi morphism for first order field theories; in
addition here the gauge-natural structure of the theories under consideration enables us to
define the {\em generalized gauge-natural Jacobi morphism} where the {\em variation vector fields} 
are Lie derivatives 
of sections of the gauge-natural bundle with respect to gauge-natural lifts. 

%-------------------------------------------------------------------
\subsection{The Bergmann--Bianchi morphism}
%--------------------------------------------------------------------

It is a well known procedure
to perform suitable integrations by
parts to decompose the conserved current $\eps$ into the sum of
a conserved current $\tilde{\eps}$ vanishing along solutions of the Euler--Lagrange equations,
the so--called {\em reduced current}, and the
formal divergence of a skew--symmetric (tensor) density $\nu$ called a {\em
superpotential} (which is then defined modulo a divergence).
Within such a procedure, the generalized Bergmann--Bianchi identities
are in fact necessary and (locally) sufficient conditions for the 
conserved current
$\epsilon$ to be not only closed but also the divergence of a 
skew-symmetric (tensor) density
along solutions of the  Euler--Lagrange equations. In \cite{PaWi03,PaWi04}, for the first time, the 
relation of the kernel of the gauge-natural Jacobi 
morphism with the kernel of the {\em 
Bergmann--Bianchi} morphism has been explicited in order 
to characterize generalized Bianchi identities in 
terms of a special class of gauge-natural lifts, namely those which have their vertical part 
in the kernel of the generalized gauge-natural Jacobi 
morphism. 

\medskip
Let now consider the term $\ome(\lam,\mathfrak{G}(\bar{\Xi})_{V})\byd -\pounds_{\bar{\Xi}} 
\rfloor \cE_{n} (\lam)$ appearing in the formulation of the First Noether Theorem given in Proposition \ref{symmetry of
L}. We stress that along sections which are not critical such a term is not vanishing, in general. In the following we
shall manipulate it to derive - under precise conditions - a strongly conserved current, \ie a current
satisfying a Noether conservation law also along non critical sections, considered for the first time by Bergmann in
\cite{Ber49}. 

In fact, as a further application of the global
decomposition formula of  vertical morphisms due to Kol\'a\v{r}
\cite{Kol83}, following essentially the procedure proposed by Bergmann in \cite{Ber49}, we can integrate by parts
$\ome(\lam,\mathfrak{G}(\bar{\Xi})_{V})$ to define the {\em generalized Bergmann--Bianchi
morphism}.

\bLm\label{Berg}
We have {\em globally}
\beq
(\pi^{4s+1}_{s+1})^{*}\ome(\lam,\mathfrak{G}(\bar{\Xi})_{V}) = 
\bet(\lam,\mathfrak{G}(\bar{\Xi})_{V}) +
F_{\ome(\lam,\mathfrak{G}(\bar{\Xi})_{V})}\,,
\eeq
where
$\bet(\lam,\mathfrak{G}(\bar{\Xi})_{V})
\equiv
E_{\ome(\lam,\mathfrak{G}(\bar{\Xi})_{V})}\,, $
and {\em locally}, $F_{\ome(\lam,\mathfrak{G}(\bar{\Xi})_{V})} = 
D_{H}M_{\ome(\lam,\mathfrak{G}(\bar{\Xi})_{V})}$.
\eLm
Coordinate expressions for the morphisms 
$\bet(\lam,\mathfrak{G}(\bar{\Xi})_{V}) $ and 
$M_{\ome(\lam,\mathfrak{G}(\bar{\Xi})_{V})}$ can be found by a 
backwards procedure (see \eg \cite{Kol83}). In particular, 
$\bet(\lam,\mathfrak{G}(\bar{\Xi})_{V})$ is nothing but the 
Euler--Lagrange morphism associated with the {\em new} Lagrangian 
$\ome(\lam,\mathfrak{G}(\bar{\Xi})_{V})$ defined on the fibered 
manifold $J_{2s}\bY_{\zet} \ucar{\bX} VJ_{2s}\cA^{(r,k)}\to \bX$.
In particular, we get the following {\em local} decomposition of 
$\ome(\lam,\mathfrak{G}(\bar{\Xi})_{V})$:
\bEq
\ome(\lam,\mathfrak{G}(\bar{\Xi})_{V}) = 
\bet(\lam,\mathfrak{G}(\bar{\Xi})_{V}) +
D_{H}\tilde{\eps}(\lam,\mathfrak{G}(\bar{\Xi})_{V}) \,,
\eEq
where we put $\tilde{\eps}(\lam,\mathfrak{G}(\bar{\Xi})_{V}) 
\equiv M_{\ome(\lam,\mathfrak{G}(\bar{\Xi})_{V})}$.

\bDf
We call the global morphism $\bet(\lam,\mathfrak{G}(\bar{\Xi})_{V}) 
\byd E_{\ome(\lam,\mathfrak{G}(\bar{\Xi})_{V})}$
the {\em generalized} {\em Bergmann--Bianchi morphism} associated
with the Lagrangian $\lam$ and the variation vector field
$\mathfrak{G}(\bar{\Xi})_{V}$.\END\eDf

As mentioned in the Introduction, the problem of the general covariance of generalized 
Bianchi identities for field theories was posed by Anderson and 
Bergmann already in $1951$ (see \cite{AnBe51}). 
This problem reflects obviously on the covariance of conserved quantities (see Remark
\ref{arbitrary1} above). Here we propose a way to deal with such open problems concerning
globality aspects. 

In fact, let now $\mathfrak{K} \byd 
\textstyle{Ker}_{\cJ(\lam,\mathfrak{G}(\bar{\Xi})_{V})}$
be the {\em kernel} of
the generalized gauge-natural morphism $\cJ(\lam,\mathfrak{G}(\bar{\Xi})_{V})$.
As a consequence of Theorem \ref{comparison} and of Lemma \ref{Berg}, we have the following covariant characterization
of the kernel  of generalized Bergmann--Bianchi morphism, the detailed proof of which will 
appear in \cite{PaWi03}.

\bTh\label{B}
The generalized Bianchi morphism is globally vanishing if and only
if $\del^{2}_{\mathfrak{G}}\lam\equiv\cJ(\lam,\mathfrak{G}(\bar{\Xi})_{V})=
0$, \ie if and only if
$\mathfrak{G}(\bar{\Xi})_{V}\in\mathfrak{K}$.
\eTh 

The gauge-natural invariance of the
variational principle {\em in its whole}  enables us to solve the 
{\em intrinsic indeterminacy}  in the conserved charges associated 
with gauge-natural symmetries of Lagrangian field theories (in 
\cite{Mat03}, for example, the special case of the gravitational 
field coupled with fermionic matter is considered and the Kosmann 
lift is then invoked as an {\em ad hoc}  choice to recover the well 
known expression of the Komar superpotential). This is 
well known to be of great  importance within the theory of Lie 
derivative of
sections of a gauge-natural bundle and notably for the Lie derivative 
of spinors (see \eg the review given in \cite{Mat03}).  As a quite strong consequence of the above Theorem, 
for any gauge-natural invariant field theory we find that the above mentioned indeterminacy can be {\em always} solved
canonically as shown by the following.

\bCr\label{fundtheorem}
Let $\lam\in \Var^{n}_{s}$ be a gauge-natural invariant generalized 
Lagrangian and let
$\mathfrak{G}(\bar{\Xi})$ be a gauge-natural lift of the principal 
infinitesimal automorphism
$\bar{\Xi}\in\cA^{r,k}$, \ie a gauge-natural symmetry of $\lam$. Then $\bar{\Xi}\in\cA^{r,k}$ is the generator of a
canonical global conserved quantity, if and only if 
$\mathfrak{G}(\bar{\Xi})_V$ satisfies the invariant condition
\beq
(-1)^{|\bsig|}D_{\bsig}\,\left(D_{\bmu}
\hat{\Xi}^{j}_{V}\left(\der_{j}(\der^{\bmu}_{i}\lam) - \sum_{|\balp |
= 0}^{s-|\bmu |}
(-1)^{|\bmu +\balp |} \frac{(\bmu +
\balp)!}{\bmu ! \balp !}
D_{\balp}\der^{\balp}_{j}(\der^{\bmu}_{i}\lam)\right)\right)=0\,.
\eeq
\eCr

In particular, the condition $j_{s}\bar{\Xi}_{V} 
=D_{\balp}(\bar{\Xi}^{i}_{V})\der^{\balp}_{i}\in \mathfrak{K}$
implies, of course,  that the components $\bar{\Xi}^i_{\balp}$
and
$\bar{\Xi}^{\gam}$ {\em are not} independent, but they are {\em 
related} in such a way
that $j_{s}\hat{\Xi}_{V}=
D_{\balp}(\hat{\Xi}^i - y^i_{\gam}\hat{\Xi}^{\gam}) \der^{\balp}_{i}$ must be a
solution of generalized gauge-natural Jacobi equations for the Lagrangian
$\lam$.

According with the above Corollary, we shall refer to {\em canonical} covariant  currents or to corresponding
superpotentials by stressing their dependence on $\mathfrak{K}$; \ie by Theorem \ref{B}
in correspondence of gauge-natural lifts satisfying  covariant Bergmann--Bianchi identities.

\bRm\label{application}
Let then  $\lam \in \Var^{n}_{s}$ be a gauge-natural Lagrangian and 
$j_{s}\hat{\Xi}_{V}\in \mathfrak{K}$ 
a gauge-natural symmetry of $\lam$. 
Being $\bet(\lam, \mathfrak{K})\equiv 0$, we have, {\em globally}, $\ome(\lam,\mathfrak{K}) = D_{H}\eps(\lam, \mathfrak{K})$,
then  from the First Noether Theorem we have
$
D_{H}(\eps(\lam, \mathfrak{K})-\tilde{\eps}(\lam, \mathfrak{K}) = 0$,
which is a so-called gauge-natural 
`strong' conservation law for the {\em global canonical} density $\eps(\lam, \mathfrak{K}) -\tilde{\eps}(\lam,
\mathfrak{K})$.

As an important application, we recall that recently the existence of 
{\em canonical gauge-natural 
superpotential} associated with $\lam$ and $\mathfrak{K}$ has been accordingly established in the framework 
of variational sequences \cite{PaWi03,PaWi04}.
In fact, let $\lam \in \Var^{n}_{s}$ be a gauge-natural Lagrangian and 
$(j_{s}\hat{\Xi},\xi)$ a gauge-natural symmetry of $\lam$. Then there exists 
a canonical global sheaf morphism 
$\nu(\lam, \mathfrak{K})$ $\in$ $\left(\Var^{n-2}_{2s-1}\right)_{\bY_{\zet} \ucar{\bX} 
\mathfrak{K}}$
such that
$
D_{H}\nu(\lam, \mathfrak{K}) = \eps(\lam, \mathfrak{K}) -\tilde{\eps}(\lam,
\mathfrak{K})$.
Notice that by the exactness of the variational sequence, the existence of a local superpotential can be 
deduced as a section of $\left(\Var^{n-2}_{2s-1}\right)_{\bY_{\zet} \ucar{\bX} 
\cA^{(r,k)}}$. This local section can be always globalized by choosing (prolongations of) principal
connections on $\bP$
\cite{FFP01}. However, such a globalization depends on the choice of the connection itself. 
Furthermore, although this choice can be always done geometrically, connections are generally the unknown to be
determined in field theories and then they {\em should not} be fixed {\em a priori} in a consistent truly covariant
field theory. Our result enables us to get global sections of the {\em reduced} sheaf 
$\left(\Var^{n-2}_{2s-1}\right)_{\bY_{\zet} \ucar{\bX} 
\mathfrak{K}}$, without fixing any connection {\em a priori}. 
\END\eRm

%-----------------------------------------------------------------------------------------------%
\subsection{Generalized symmetries and Bergmann--Bianchi identities}
%-----------------------------------------------------------------------------------------------%

As well known, the Second Noether Theorem deals 
with invariance properties of the Euler-Lagrange equations (so-called 
generalized symmetries or also Bessel-Hagen symmetries, see \eg the 
fundamental papers \cite{Tra67}). Although symmetries of a Lagrangian 
turn out to be also symmetries of the Euler--Lagrange morphism the 
converse is not true, in general.

In particular, although for a 
gauge-natural invariant Lagrangian $\lam$ we always have 
$\cL_{j_{s}\bar{\Xi}}\lam=0$,   $\cL_{j_{s}\bar{\Xi}_V}\lam$  does 
not need to be zero in principle; however when the second variation 
$\del^{2}_{\mathfrak{G}}\lam$ is required to vanish then 
$\cL_{j_{s}\bar{\Xi}_V} \cE_{n}(\lam)$ surely vanishes, \ie 
$j_{s}\bar{\Xi}_V$ is a generalized or Bessel--Hagen symmetry. The 
symmetries of the Euler--Lagrange morphism (Second Noether Theorem) 
impose some constraints on the conserved quantities associated with 
gauge-natural symmetries of $\lam$ (see \eg \cite{AnBe51}).

Symmetries of the Euler--Lagrange morphism are clearly related with invariance properties of 
$\ome(\lam,\mathfrak{G}(\bar{\Xi})_{V})\byd -\pounds_{\bar{\Xi}} 
\rfloor \cE_{n} (\lam)$. We stress that, because of linearity properties of $\pounds$,
$\ome(\lam,\mathfrak{G}(\bar{\Xi})_{V})$ can be considered as a new Lagrangian, defined on an extended space; thus
Theorems
\ref{comparison} and
\ref{B} can provide us with some kind of Noether conservation law associated with the induced invariance properties of
$\ome(\lam,\mathfrak{G}(\bar{\Xi})_{V})$.

First of all let us make the following important consideration.

\bPr\label{Pr3}
For each $\bar{\Xi}\in \cA^{(r,k)}$ such that $\bar{\Xi}_{V}\in \mathfrak{K}$, we have
\bEq
\cL_{j_{s}\bar{\Xi}_{H}}\ome(\lam,
\mathfrak{K})=-D_{H}(-j_{s}\pounds_{\bar{\Xi}_{V}} 
\rfloor p_{D_{V}\ome(\lam,\mathfrak{K})}) \,.
\eEq
\ePr

\bPf
The horizontal splitting gives us $\cL_{j_{s}\bar{\Xi}}\ome(\lam,\mathfrak{K})=\cL_{j_{s}\bar{\Xi}_{H}}\ome(\lam,\mathfrak{K}) +
\cL_{j_{s}\bar{\Xi}_{V}}\ome(\lam,\mathfrak{K})$. Furthermore, $\ome(\lam,\mathfrak{K})\equiv - \pounds_{\bar{\Xi}} \rfloor
\cE_{n}(\lam) = \cL_{j_{s}\bar{\Xi}}\lam - d_{H}(-j_{s}\pounds_{\bar{\Xi}} 
\rfloor p_{d_{V}\lam}+ \xi \rfloor \lam)$; so that
$
\cL_{j_{s}\bar{\Xi}_{V}}\ome(\lam,\mathfrak{K})=\cL_{j_{s}\bar{\Xi}_{V}}\cL_{j_{s}\bar{\Xi}}\lam=\cL_{j_{s}[\bar{\Xi}_{V},
\bar{\Xi}_{H}]}\lam$. On the other hand we have 
$
\cL_{j_{s}\bar{\Xi}_{H}}\ome(\lam,\mathfrak{K}) = \cL_{j_{s}[\bar{\Xi}_{H},
\bar{\Xi}_{V}]}\lam=-\cL_{j_{s}\bar{\Xi}_{V}}\ome(\lam,\mathfrak{K})$.

Recall now that $\bar{\Xi}_{V}\in
\mathfrak{K}$ if and only if $\bet(\lam,\mathfrak{K})=0$. 
Since 
\beq
&& \cL_{j_{s}\bar{\Xi}_{V}}\ome(\lam,\mathfrak{K})=
 - \pounds_{\bar{\Xi}_{V}} \rfloor \cE_{n}(\ome(\lam,\mathfrak{K})) 
+ D_{H}(-j_{s}\pounds_{\bar{\Xi}_{V}} 
\rfloor p_{D_{V}\ome(\lam,\mathfrak{K})}) = \\
& & = \bet(\lam,\mathfrak{K})+D_{H}(-j_{s}\pounds_{\bar{\Xi}_{V}} 
\rfloor p_{D_{V}\ome(\lam,\mathfrak{K})})\,,
\eeq
 we get the assertion.
\END
\ePf

It is easy to realize that, because of the gauge-natural invariance of the generalized
Lagrangian $\lam$, the new generalized Lagrangian $\ome(\lam,\mathfrak{K})$ 
is gauge-natural invariant too, \ie
$\cL_{j_{s}\bar{\Xi}}\,\ome(\lam,
\mathfrak{K})=0$.
However, a stronger result holds true. In fact, we can state the following naturality property for
$\ome(\lam,\mathfrak{K})$, which provides some more information concerning the Hamiltonian structure of gauge-natural
field theories \cite{FPW04}.

\bPr\label{Pr4}
Let $\bar{\Xi}_{V}\in
\mathfrak{K}$. We have
$
\cL_{j_{s}\bar{\Xi}_{H}}\ome(\lam,
\mathfrak{K})=0$.
\ePr 

\bPf
In fact, when $\bar{\Xi}_{V}\in
\mathfrak{K}$, by the theory of iterated Lie derivatives of sections \cite{KMS93}, we have $\cL_{j_{s}\bar{\Xi}_V}\ome(\lam,
\mathfrak{K})=[\bar{\Xi}_V,\bar{\Xi}_V]\rfloor \cE_n (\lam)+ 
\bar{\Xi}_V \rfloor\cL_{j_{s}\bar{\Xi}_V}\cE_n (\lam) =0$. Thus
$\cL_{j_{s}\bar{\Xi}}\ome(\lam,
\mathfrak{K})= \cL_{j_{s}\bar{\Xi}_V}\ome(\lam,
\mathfrak{K})+\cL_{j_{s}\bar{\Xi}_H}\ome(\lam,
\mathfrak{K})= \cL_{j_{s}\bar{\Xi}_H}\ome(\lam,
\mathfrak{K})=0$.
\QED
\ePf

As a consequence of Propositions \ref{Pr3} and \ref{Pr4}, corresponding to any $\mathfrak{G}(\bar{\Xi})_{H}$, we get the
existence of a generalized Noether conserved current (which could be interpreted as a {\em generalized energy--momentum
tensor} for
$\ome(\lam,
\mathfrak{K})$).

\bCr
Let $\bar{\Xi}_{V}\in
\mathfrak{K}$. We have the {\em covariant} conservation law
\bEq
D_{H}(-j_{s}\pounds_{\bar{\Xi}_{V}} 
\rfloor p_{D_{V}\ome(\lam,\mathfrak{K})})=0\,.
\eEq
\eCr

%-----------------------------------------------------------------------------%
\subsection*{Acknowledgments}
%-----------------------------------------------------------------------------%

Thanks are due to Prof. I. Kol\'a\v{r} for helpful discussions and in particular for having brought
to our attention the role played by the theory of iterated Lie derivatives of sections of fibered
bundles.

%------------------------------------------------------------------------%

\end{document}